\documentclass[11pt]{article}
\pdfoutput=1
\usepackage{fullpage}
\usepackage{cite}
\usepackage{amsmath}
\usepackage{amssymb}
\usepackage{graphicx}

\title{}
\date{}
\def\para{\\ [-2mm]}

\def\eqn#1{eq.~(\ref{#1})} \def\Eqn#1{Equation~(\ref{#1})}
\def\eqns#1#2{eqs.~(\ref{#1}) and~(\ref{#2})}

\def \be  {\begin{equation}}
\def \ee  {\end{equation}}
\def \ba  {\begin{eqnarray}}
\def \ea  {\end{eqnarray}}
\newcommand{\nn}{\nonumber}

\def\ie{{i.e.,~}}
\def\viz{{viz.,~}}
\def\eg{{e.g.,~}}

\def\id{  {\mathsf{1}\kern -3pt \mathsf{l} } }

\def\IZ{\relax\ifmmode\mathchoice
{\hbox{\cmss Z\kern-.4em Z}}{\hbox{\cmss Z\kern-.4em Z}}
{\lower.4pt\hbox{\cmsss Z\kern-.4em Z}}
{\lower1.2pt\hbox{\cmsss Z\kern-.4em Z}}\else{\cmss Z\kern-.4em Z}\fi}
\newcommand{\Z}{\mathsf{Z}\kern -5pt \mathsf{Z}}
\newcommand{\unit}{\mathsf{1}\kern -3pt \mathsf{l}}

\def\barpsi{{\bar\psi}}
\def\X { {\barpsi } }
\def\Y { {\psi} }

\def\Sai{ S_{a,i} }

\def\ve {\varepsilon}
\def\ti {\textsf{i}}
\def\tj {\textsf{j}}

\def\ta {\textsf{a}}
\def\tb {\textsf{b}}
\def\tc {\textsf{c}}
\def\td {\textsf{d}}
\def\te {\textsf{e}}

\def\cA {  {\cal A} }

\def\cO {  {\cal O} }

\def\bc{  {\bf c} }

\def\sumr{  \sum_{r=1}^3 }

\begin{document}

\titlepage
\begin{flushright}
BOW-PH-163\\
MCTP-16-18\\
\end{flushright}

\vspace{3mm}

\begin{center}

{\Large\bf\sf
Color-factor symmetry and BCJ relations for QCD amplitudes
}

\vskip 1.5cm

{\sc
Robert W. Brown$^a$
and Stephen G. Naculich$^{b,c}$
}

\vskip 0.5cm
$^a${\it
Department of Physics\\
Case Western Reserve University\\
Cleveland, OH 44106 USA
}

\vskip 0.5cm
$^b${\it
Department of Physics\\
Bowdoin College\\
Brunswick, ME 04011 USA
}

\vskip 0.5cm
$^c${\it
Michigan Center for Theoretical Physics (MCTP)\\
Department of Physics\\
University of Michigan\\
Ann Arbor, MI 48109 USA
}

\vspace{5mm}
{\tt
rwb@case.edu,  naculich@bowdoin.edu
}
\end{center}

\vskip 1.5cm

\begin{abstract}

Tree-level $n$-point gauge-theory amplitudes
with $n-2k$ gluons and 
$k$ pairs of (massless or massive) particles 
in the fundamental (or other) representation 
of the gauge group
are invariant under a set of symmetries 
that act as momentum-dependent shifts on the color factors
in the cubic decomposition of the amplitude.
These symmetries lead to gauge-invariant constraints on 
the kinematic numerators.
They also directly imply the BCJ relations 
among the Melia-basis primitive amplitudes
previously obtained by Johansson and Ochirov.
\end{abstract}

\vspace*{0.5cm}

\vfil\break

\section{Introduction}
\setcounter{equation}{0}

Bern, Carrasco, and Johansson discovered that the 
color-ordered amplitudes of tree-level $n$-gluon amplitudes
obey a set of helicity-independent relations,
linear relations whose coefficients depend only on 
Lorentz-invariant combinations of the momenta of the 
particles \cite{Bern:2008qj}.
They inferred these relations by imposing color-kinematic 
duality on the kinematic numerators 
appearing in a cubic decomposition of the amplitude.
These BCJ relations have been proven through a variety of approaches:
string theory \cite{BjerrumBohr:2009rd,Stieberger:2009hq},
on-shell BCFW recursion \cite{Feng:2010my,Chen:2011jxa},
the Cachazo-He-Yuan 
representation \cite{Cachazo:2012uq,Cachazo:2013iaa,Cachazo:2013iea},
and most recently by using the invariance of the
$n$-gluon amplitude under a momentum-dependent shift 
of the color factors appearing in the cubic 
decomposition \cite{Brown:2016mrh}.
\para

Several authors have turned their attention 
to the question of color-kinematic duality in gauge-theory amplitudes 
containing (massless or massive) quarks or other particles 
in the fundamental representation of the gauge 
group \cite{Johansson:2014zca,Naculich:2014naa,Weinzierl:2014ava,Johansson:2015oia,Mastrolia:2015maa,delaCruz:2015dpa,delaCruz:2015raa,Chiodaroli:2015rdg,He:2016dol}. 
Melia characterized an independent basis of primitive amplitudes 
for tree-level amplitudes that involve gluons 
and an arbitrary number of pairs of differently flavored 
fundamentals \cite{Melia:2013bta,Melia:2013epa,Melia:2015ika}.
By imposing color-kinematic duality, Johansson and Ochirov
showed that these Melia amplitudes satisfy a set of 
BCJ relations that are formally identical\footnote{
\ie when written in terms of $k_a \cdot k_b$, where $k_a$
is the momentum of one of the gluons}
to a subset of the $n$-gluon BCJ relations \cite{Johansson:2015oia}.
The same relations had previously been established,
also by assuming color-kinematic duality,
for amplitudes with gluons and a single pair of massive 
fundamentals \cite{Naculich:2014naa}.
These relations were subsequently proven using BCFW on-shell recursion 
by de la Cruz, Kniss, and Weinzierl \cite{delaCruz:2015dpa}.
\para

In this paper, we demonstrate that all the BCJ relations found
in ref.~\cite{Johansson:2015oia}
are a consequence of the color-factor symmetry possessed 
by the amplitude.
This symmetry, recently established in ref.~\cite{Brown:2016mrh},
states that the amplitude is invariant 
under certain momentum-dependent shifts of the color factors.
We show that for the $n$-point amplitude $\cA_{n,k}$,
with $k$ pairs of fundamentals and $n-2k$ gluons, 
there is an $(n-3)!/k!$-parameter family 
of color-factor shifts associated with 
each external gluon in the amplitude.
The total number of independent color-factor shifts
is given by $(n-2k) (n-3)!/k!$ for $k\ge 2$,
and $(n-3) (n-3)!$ for $k=0$ and $1$.
We show that the color factors 
introduced by Johansson and Ochirov \cite{Johansson:2015oia} 
transform in a particularly simple way under a color-factor shift.
Consequently, the BCJ relations 
among the primitive amplitudes follow immediately from the 
invariance under color-factor shifts of the 
Melia-Johansson-Ochirov proper decomposition of the amplitude.
\para

Johansson and Ochirov found that the BCJ relations 
satisfied by the Melia primitives
are directly tied to the presence of external gluons
in the amplitude.  
Each set of fundamental relations corresponds 
to one of the external gluons,
and when the amplitude contains no external gluons,
no relations exist among the primitive amplitudes.
The link shown in this paper 
between the color-factor symmetry and BCJ relations
makes sense of this result, 
since the color-factor shifts are associated 
with external gluons in the amplitude.
The number of independent BCJ relations
is precisely equal to the dimension of the color-factor group,
and the absence of color-factor symmetry for amplitudes
with no gluons explains the absence of relations
among the primitive amplitudes.
\para

We also describe the cubic vertex expansion for 
$\cA_{n,k}$ that was introduced in ref.~\cite{Brown:2016mrh},
and show that the color-factor symmetry leads to 
gauge-invariant constraints on the kinematic numerators.
These constraints are less stringent than
the kinematic Jacobi relations
but are nonetheless sufficient to imply the validity
of the BCJ relations.
\para

The contents of this paper are as follows.
In sec.~\ref{sec:proper},  
we describe the Melia basis of primitive amplitudes 
for $\cA_{n,k}$ and the corresponding color factors
found by Johansson and Ochirov.
In sec.~\ref{sec:bcj},  
we review the BCJ relations for the 
primitive amplitudes of $\cA_{n,k}$.
In sec.~\ref{sec:cfs}, 
we describe the color-factor symmetry 
and determine the shifts of the Johansson-Ochirov color factors.
In sec.~\ref{sec:bcjcfs}, we review the proof of the 
invariance of $\cA_{n,k}$ under the color-factor shift,
and show that the BCJ relations 
as well as constraints on kinematic numerators
follow directly from this invariance.   
Section~\ref{sec:concl} contains our conclusions.

\section{Proper decompositions of gauge-theory amplitudes}
\setcounter{equation}{0}
\label{sec:proper}

In this section,  we describe proper decompositions of gauge-theory
amplitudes with particles in the adjoint and fundamental representations;
that is, decompositions in terms of an independent set of color factors
and gauge-invariant primitive amplitudes.
These include the Del Duca-Dixon-Maltoni decomposition
of $n$-gluon amplitudes and the recently developed Melia-Johansson-Ochirov
decomposition of amplitudes with $n-2k$ gluons and $k$ pairs of 
fundamentals.
\para

Consider a tree-level $n$-point gauge-theory amplitude $\cA_{n,k}$
with $n-2k$ gluons and 
$k$ pairs of (massless or massive)
particles $\psi$ and $\barpsi$
in the fundamental (and antifundamental) 
representation\footnote{The particles $\psi$ can actually be in any 
representation, but we refer to the fundamental for convenience.}
of the gauge group,
with arbitrary spin $\le 1$.
This amplitude may be written in a 
cubic decomposition\cite{Bern:2008qj}
\be
\cA_{n,k} ~=~ \sum_i {c_i ~ n_i \over d_i }
\label{cubicdecomp}
\ee
namely, as a sum over diagrams $i$ consisting only of trivalent vertices.
The color factor $c_i$ associated with each diagram
is obtained by sewing together $ggg$ vertices $f_{\ta \tb \tc} $
and $\barpsi g \psi $ vertices $(T^\ta )^{\ti}_{~\tj}$,
where $T^\ta$ denote generators in the fundamental (or other) representation.
The kinematic numerator $n_i$ carries information about the spin state 
of the particles.
The denominator $d_i$ consists of the product of the inverse propagators
associated with the diagram.
Contributions from Feynman diagrams with quartic vertices 
(either $gggg$ or $\barpsi g g \psi $ in the case of a 
scalar or vector $\psi$)
are parceled out among the cubic diagrams.
\para

In general, the color factors $c_i$ are not independent,
but obey a set of Jacobi relations of the form
\be
c_i + c_j + c_k ~=~ 0  
\ee
by virtue of the group theory identities
$ f_{\ta \tb \te} f_{\tc \td \te} 
+f_{\ta \tc \te} f_{\td \tb \te} 
+f_{\ta \td \te} f_{\tb \tc \te}=0 $
and 
$ f_{\ta \tb \tc} ( T^{\tc} )^{\ti}_{~\tj} 
= [ T^{\ta}, T^{\tb} ]^{\ti}_{~\tj}  $.
Because of this, the kinematic numerators $n_i$ are not well-defined,
but may undergo generalized gauge 
transformations \cite{Bern:2010ue,Bern:2010yg}
that leave \eqn{cubicdecomp} unchanged.
In principle, the Jacobi relations for the color factors may be solved
in terms of an independent set of color factors $C_j$,
and the amplitude written in a {\it proper} color 
decomposition \cite{Melia:2015ika}
\be
\cA_{n,k} ~=~ \sum_j   C_j A_j \,.
\label{proper}
\ee
The coefficients $A_j$, referred to as primitive amplitudes,
receive contributions from several 
different terms in \eqn{cubicdecomp}.
Because the $C_j$ are independent,
the primitive amplitudes will be well-defined (gauge invariant).
\para

The primitive amplitudes may be chosen to be an independent subset 
of color-ordered amplitudes.
The color-ordered amplitude $A(\alpha)$ 
is computed,
using color-ordered Feynman rules \cite{Dixon:1996wi},
from the sum of planar Feynman diagrams
whose external legs are in the order specified by the permutation $\alpha$.
The color-ordered amplitudes are not independent, 
but obey various relations such as cyclicity 
and reflection invariance \cite{Mangano:1990by}
and the Kleiss-Kuijf relations \cite{Kleiss:1988ne}.
For pure gluon amplitudes ($k=0$), 
an independent set of $(n-2)!$ color-ordered amplitudes are those
belonging to the Kleiss-Kuijf basis 
$A(1, \gamma(2),  \cdots, \gamma(n-1), n ) $,
in which the positions of two of the gluons are fixed
and $\gamma$ is a permutation of $\{2, \cdots, n-1\}$.
The corresponding color factors $C_j$
are the $(n-2)!$ half-ladder diagrams 
\be
\bc_{1 \gamma n} 
~=~  
\sum_{\tb_1,\ldots,\tb_{n{-}3}} 
f_{\ta_{1} \ta_{\gamma(2)} \tb_1}
f_{\tb_1 \ta_{\gamma(3)} \tb_2}
\cdots f_{\tb_{n{-}3} \ta_{\gamma(n{-}1)} \ta_n} \,, 
\qquad
\gamma \in S_{n-2} 
\ee
and the proper decomposition 
\be
\cA_{n,0} (g_1, g_2,  \cdots, g_{n})
~=~ \sum_{\gamma \in S_{n-2}}  
\bc_{1 \gamma n } \, A(1, \gamma(2), \cdots, \gamma(n-1), n) 
\label{ddm}
\ee
is known as the Del Duca-Dixon-Maltoni decomposition \cite{DelDuca:1999rs}.
\para

The $n$-point gauge-theory amplitude with $n-2$ gluons
and one pair of fundamentals has a similar proper 
decomposition  \cite{Kosower:1987ic,Mangano:1988kk}
\be
\cA_{n,1} (\barpsi_1, \psi_2, g_3, \cdots, g_{n})
~=~ \sum_{\gamma \in S_{n-2}}  
C_{1\gamma 2} \, A (1, \gamma(3), \cdots, \gamma(n),2)
\label{ddmfund}
\ee
where the independent color factors are given by 
\be
C_{1\gamma 2} 
~=~
\left( {T}^{\ta_{\gamma(3)}}{T}^{\ta_{\gamma(4)}}
\cdots {T}^{\ta_{\gamma(n)}} \right)^{\ti_1}_{~~ \ti_2} 
\label{halfladderfund}
\qquad {\rm ~for~} k=1
\ee
with $\gamma$ a permutation of $\{3, \cdots, n\}$.
\para

Finding a proper decomposition for amplitudes with more 
than one pair of fundamentals is a more subtle problem,
but was recently solved by Melia \cite{Melia:2013bta,Melia:2013epa}
and Johansson and Ochirov \cite{Johansson:2015oia}.
Consider an $n$-point amplitude of the form 
\be
\cA_{n,k} ( 
\barpsi_1, \psi_2, \barpsi_3, \psi_4, \cdots, \barpsi_{2k-1}, \psi_{2k},
g_{2k+1}, \cdots, g_n)
\ee
where particles $\psi$ in the fundamental representation
have even labels, 
and particles $\barpsi$ in the antifundamental representation 
have odd labels.\footnote{In 
refs.~\cite{Melia:2013bta,Melia:2013epa,Johansson:2015oia},
$\psi$ and $\barpsi$ are referred to as quarks and antiquarks,
but they could just as easily be scalar or vector particles.}
We assume that the $\psi_{2\ell}$ all have different flavors 
(and possibly different masses), 
with $\barpsi_{2\ell-1}$ having the corresponding antiflavor 
(and equal mass) to $\psi_{2\ell}$.
This assumption entails no loss of generality since amplitudes with 
multiple pairs of fundamentals with the same flavor and mass 
can be obtained by setting flavors and masses equal 
and summing over permutations.
As shown in ref.~\cite{Melia:2013bta},
many of the color-ordered amplitudes $A(\alpha)$ associated 
with this amplitude vanish because 
no planar Feynman diagram with external legs in the order $\alpha$ 
can be drawn that does not violate flavor conservation. 
Moreover,
the non-vanishing color-ordered amplitudes
satisfy further relations in addition to the Kleiss-Kuijf relations.
Melia identified a subset of $(n-2)!/k!$ 
color-ordered amplitudes that form an independent set. 
\para

To describe the Melia basis of primitive amplitudes,
we must recall the definition of a Dyck word \cite{Melia:2013bta}.
A Dyck word of length $2r$ is a string composed of  
$r$ letters $\X$ and $r$ letters $\Y$
such that the number of $\X$'s preceding any point in the string
is greater than the number of preceding $\Y$'s.
An easy way to understand this is to 
visualize $\X$ as a left bracket $\{$ 
and $\Y$ as a right bracket $\}$, 
in which case a Dyck word corresponds to a well-formed set of brackets.
The number of such words is $(2r)!/(r+1)!r!$, the $r$th Catalan number. 
For example 
for $r=1$ there is only one Dyck word: $\{ \} $, 
for $r=2$ there are two: $\{\} \{\} $ and $\{\{\} \} $, 
and for $r=3$ there are five:
$\{\} \{\} \{\} $, 
$\{\} \{\{\} \} $, 
$\{\{\} \} \{\} $,
$\{\{\} \{\} \} $, and
$\{\{\{\} \} \} $. 
\para

The Melia basis is the set of color-ordered amplitudes
$ A( 1, \gamma(3), \cdots, \gamma(n), 2 )$,
where $\gamma$ is any permutation of 
$\{ 3, \cdots, n  \}$ 
such that the set of $k-1$ $\X$ and $k-1$ $\Y$ 
in $\gamma$ form a Dyck word of length $2k-2$.
The gluons may be distributed anywhere 
among the $\barpsi$ and $\psi$ in $\gamma$. 
The  number of distinct allowed patterns of 
$\barpsi$, $\psi$, and $g$ is given by the number of Dyck words
of length $2k-2$ times the number of ways of distributing
$n-2k$ gluons among the letters of the Dyck word
\be 
{(2k-2)! \over k!(k-1)! }
\times 
{n-2 \choose  2k-2}  \,.
\ee
For each allowed pattern, 
there are $(n-2k)!$ distinct 
choices for the gluon labels,
and $(k-1)!$ choices for the $\barpsi$ labels.
The label on each $\psi$ is then fixed: 
it must have the flavor of the nearest unpaired $\barpsi$ to its left.
Thus, for example,
for $\cA_{6,3}$ the allowed permutations $\gamma$ are
$ \barpsi_3 \psi_4 \barpsi_5 \psi_6 $,
$ \barpsi_5 \psi_6 \barpsi_3 \psi_4 $,
$ \barpsi_3 \barpsi_5 \psi_6 \psi_4 $, and
$ \barpsi_5 \barpsi_3 \psi_4 \psi_6 $,
whereas for $\cA_{5,2}$ the allowed permutations are
$ \barpsi_3 \psi_4 g_5$,
$ \barpsi_3 g_5 \psi_4 $, and 
$ g_5\barpsi_3 \psi_4 $.
The multiplicity of the Melia basis is given by 
\be
{(2k-2)! \over k!(k-1)! }
\times 
{n-2 \choose  2k-2} 
\times 
(n-2k)! 
\times (k-1)! 
~=~ {(n-2)!  \over k!}
\ee
as found in ref.~\cite{Melia:2013epa}.
\para

\begin{figure}
\begin{center}
\includegraphics[scale=1.0,trim=100 690 80 50,clip=true]{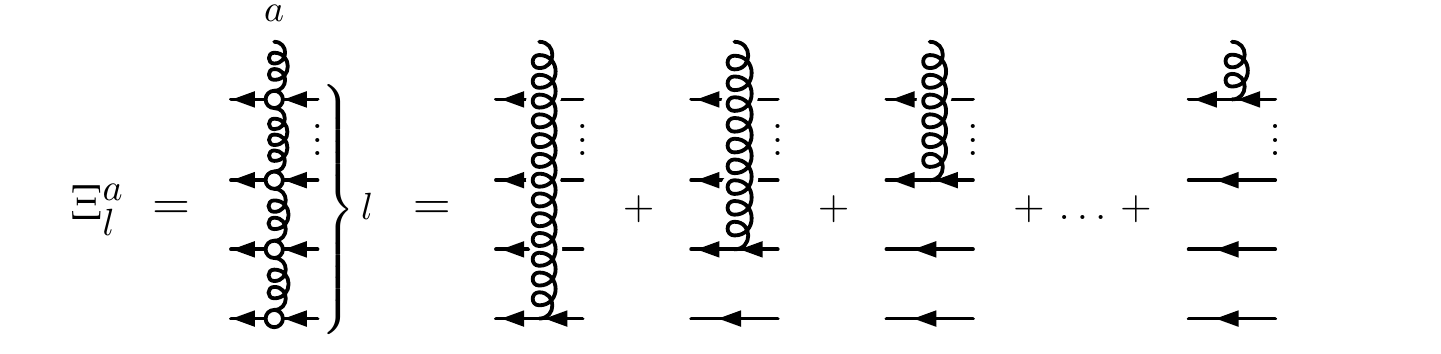}
\caption{Diagrammatic form of the operator $\Xi^\ta_l$.}
\label{fig:u1}
\end{center}
\end{figure}

Having chosen the color-ordered amplitudes in the Melia basis 
$A(1, \gamma, 2)$ 
to be the primitive amplitudes 
in a proper decomposition (\ref{proper}),
one may ask what is the corresponding set 
of independent color factors $C_{1 \gamma 2}$?
Johansson and Ochirov (JO) posed and solved this problem 
in ref.~\cite{Johansson:2015oia}.
For $k=1$, 
all permutations $\gamma$ of $\{ 3, \cdots, n\}$ are allowed, 
and the corresponding color factors are given by \eqn{halfladderfund},
which we can conveniently denote as 
$\{1| \gamma(3) \cdots \gamma(n) | 2\}$.
For $k>1$,  the JO color factors  $C_{1 \gamma 2}$
are given by linear combinations of cubic color factors $c_i$.
To obtain these linear combinations, 
one starts with $\{1| \gamma(3) \cdots \gamma(n) | 2\}$
and then replaces each $\barpsi_a$ appearing in $\gamma$
with the expression
$\{ a | \, T^\tb\!\otimes \Xi_{l-1}^\tb$,
each $\psi_a$ with the expression
$ | a \} $, 
and each gluon $g_a$ with the operator 
$\Xi_{l}^{\ta_a}$,
where 
\be
\Xi_l^\ta ~=~
\sum_{s=1}^{l}
\underbrace{1 \otimes \cdots \otimes 1 \otimes
\overbrace{T^\ta \otimes 1 \otimes \cdots \otimes 1}^{s}
}_{l}  \,.
\label{Xi}
\ee
The integer $l$ denotes the level of bracket ``nestedness'' 
at the point where $\Xi_l^{\ta_a}$ is inserted,
\ie the number of left brackets 
minus the number of right brackets to the left of the operator.
The operator $\Xi_{l}^\ta$ is a tensor product of $l$ copies 
of the Lie algebra,
where the sum runs over each position $s$ in the tensor product.
Each copy of the Lie algebra representation corresponds
to a particular nestedness level, 
starting from level $l$ (the leftmost copy) 
down to level one (the rightmost copy).
The $\{a|$ and $|a\}$ act only on the copy of the Lie algebra 
at their corresponding nestedness level $l$.
The operator $\Xi_l^\ta$ is 
conveniently represented by fig.~\ref{fig:u1},
in which the open circles represent summation over the possible locations 
where the gluon line can attach.
For $k=1$, 
$\Xi_{1}^\ta = T^\ta$ so the JO prescription  
simply reduces to \eqn{halfladderfund}.
For $k>1$, we end up with (a linear combination of) 
$k$ strings of generators of the form (\ref{halfladderfund}).
For a more detailed description and justification of the JO color factors
and many illuminating examples, 
we refer the reader to ref.~\cite{Johansson:2015oia}.
\para

Having defined the primitive amplitudes and corresponding color factors,
we can write the proper (Melia-Johansson-Ochirov) decomposition 
of the amplitude $\cA_{n,k}$ as
\be
\cA_{n,k} ( 
\barpsi_1, \psi_2, \barpsi_3, \psi_4, \cdots, \barpsi_{2k-1}, \psi_{2k},
g_{2k+1}, \cdots, g_n)
~=~ \sum_{\gamma \in \text{Melia basis}} 
C_{1\gamma 2}  \,  A (1, \gamma(3), \cdots, \gamma(n), 2) \,.
\label{mjo}
\ee
This expression is equivalent to \eqn{cubicdecomp},
as was explicitly verified for $n \le 8$ 
(and for $n=9$, $k=4$) 
in ref.~\cite{Johansson:2015oia},
and was proven for all $n$ in ref.~\cite{Melia:2015ika}.
Noting the similarity to \eqn{ddmfund},
the JO color factors $C_{1\gamma 2}$ 
can be considered the natural generalization 
of the half-ladder color factors \eqn{halfladderfund}.
We will see more evidence of this correspondence in sec.~\ref{sec:cfs}.
\para

\begin{figure}[b]
\begin{center}
\includegraphics[scale=1.0,trim=80 710 50 65,clip=true]{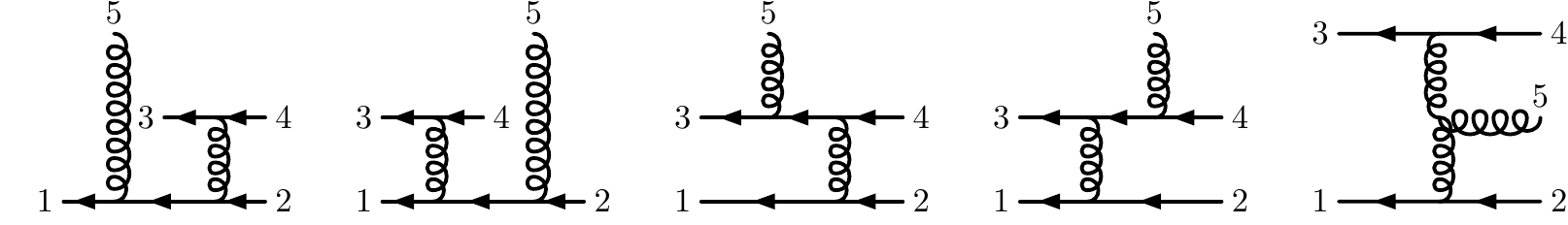}
\caption{\small Color factors $c_1$ through $c_5$ for $\cA_{5,2}$.}
\label{fig:u2}
\end{center}
\end{figure}

\begin{figure}
\begin{center}
\includegraphics[scale=1.0,trim=150 705 150 80,clip=true]{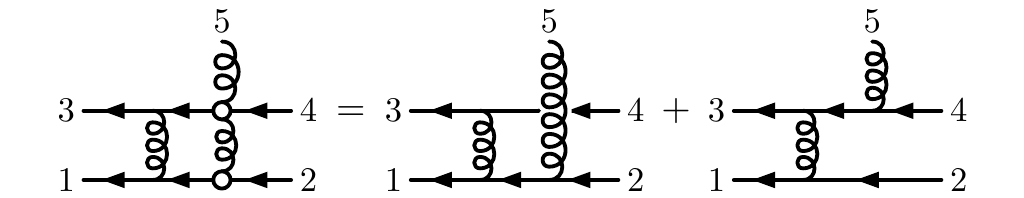}
\caption{Color factor $C_{13542}$ for the amplitude $\cA_{5,2} $.}
\label{fig:u3}
\end{center}
\end{figure}

We conclude this section with a concrete example of the 
ideas discussed above, the five-point amplitude
with one gluon and two pairs of fundamentals
$ \cA_{5,2} ( \barpsi_1, \psi_2, \barpsi_3, \psi_4, g_5)$.
Five cubic diagrams contribute to this amplitude
(see fig.~\ref{fig:u2})
\be
\cA_{5,2} ~=~ \sum_{i=1}^5 {c_i ~ n_i \over d_i } 
\label{fivepointcubicdecomp}
\ee
where the color factors and denominators have the 
form \cite{Johansson:2015oia}
\begin{align}
  c_1 &= (T^{\ta_5} T^\tb )^{\ti_1}_{~\ti_2} (T^\tb)^{\ti_3}_{~\ti_4} \,, \qquad~
& d_1 &= (s_{15}\!-\!m_1^2) s_{34} = 2 s_{34} \,  k_1 \cdot k_5 \, \,, \nn\\
  c_2 &= (T^\tb T^{\ta_5})^{\ti_1}_{~\ti_2} (T^\tb)^{\ti_3}_{~\ti_4} \,, \qquad~
& d_2 &= (s_{25}\!-\!m_1^2) s_{34} = 2 s_{34} \,  k_2 \cdot k_5 \, \,, \nn\\
  c_3 &= (T^\tb )^{\ti_1}_{~\ti_2} (T^{\ta_5} T^\tb)^{\ti_3}_{~\ti_4} \,, \qquad~
& d_3 &= s_{12} (s_{35}\!-\!m_3^2) = 2 s_{12} \, k_3 \cdot k_5 \, \,, \nn\\
  c_4 &= (T^\tb )^{\ti_1}_{~\ti_2} (T^\tb T^{\ta_5})^{\ti_3}_{~\ti_4} \,, \qquad~
& d_4 &= s_{12} (s_{45}\!-\!m_3^2) = 2 s_{12} \, k_4 \cdot k_5 \, \,, \nn\\
  c_5 &= f^{\ta_5 \tb \tc}\;\!  (T^\tb)^{\ti_1}_{~\ti_2} (T^\tc)^{\ti_3}_{~\ti_4} \,, \qquad~
& d_5 &= s_{12} s_{34} 
\label{fivepointcolorfactors}
\end{align} 
with $s_{ij} = (k_i + k_j)^2$.
The kinematic numerators $n_i$ depend on the spin of the fundamentals;
explicit expressions for spin one-half fundamentals
are given in ref.~\cite{Johansson:2015oia}.
By virtue of
$[T^{\ta},T^{\tb}]^{\ti}_{~\tj} = f_{\ta \tb \tc} ( T^{\tc} )^{\ti}_{~\tj} $,
the color factors obey two Jacobi relations
\be
   c_1 - c_2 + c_5 ~=~ 0 \,, \qquad \qquad
   c_3 - c_4 - c_5 ~=~ 0 \,.
\label{fivepointjacobi} 
\ee
Thus the proper decomposition contains three terms.
The Melia basis primitive amplitudes are
\be
   A(1,5,3,4,2)
 ~=~ \frac{n_1}{d_1} - \frac{n_3}{d_3} - \frac{n_5}{d_5} \,, 
\qquad
   A(1,3,5,4,2)
 ~=~ \frac{n_3}{d_3} + \frac{n_4}{d_4} \,, 
\qquad 
   A(1,3,4,5,2)
~=~ \frac{n_2}{d_2} - \frac{n_4}{d_4} + \frac{n_5}{d_5} 
\label{primitives5point}
\ee
and the corresponding JO color factors are 
\begin{align}
   C_{15342} ~&=~ 
\{1|  T^{\ta_5} \{3| T^\tb \otimes T^\tb  |4\} |2\}  
~=~c_1 \,,
\nn\\
   C_{13542} ~&=~ 
\{1|  \{3 |  (T^\tb \otimes T^\tb) \Xi^{a_5}_2 |4\} |2\}  
~=~ c_2 + c_4 \,,
\nn\\
   C_{13452} ~&=~ 
\{1|  \{3| T^\tb  \otimes T^\tb |4\} T^{\ta_5} |2\}  
~=~ c_2 
\label{fivepointJO}
\end{align}
where $C_{13542}$ is represented graphically in fig.~\ref{fig:u3}.
It is straightforward to check that the proper decomposition 
\be
\cA_{5,2}  ~=~ 
C_{15342} \, A(1,5,3,4,2) ~+~C_{13542} \, A(1,3,5,4,2) ~+~C_{13452} \, A(1,3,4,5,2)
\ee
is equal to \eqn{fivepointcubicdecomp}.
\para

\section{BCJ relations for QCD amplitudes}
\setcounter{equation}{0}
\label{sec:bcj}

In refs.~\cite{Johansson:2014zca,Johansson:2015oia},
Johansson and Ochirov explored whether tree-level QCD amplitudes 
with both gluons and quarks obey color-kinematic duality; 
that is,
whether there exists a generalized gauge 
in which the kinematic numerators $n_i$ 
in the cubic decomposition (\ref{cubicdecomp})
satisfy the same algebraic relations 
as do the color factors $c_i$
for amplitudes with particles 
in the adjoint and fundamental representations
(see also ref.~\cite{Mastrolia:2015maa}).
Color-kinematic duality is well-established for pure gluon 
amplitudes \cite{Bern:2008qj,Bern:2010yg,Bern:2013yya},
for massless particles in supersymmetric Yang-Mills multiplets that contain 
gluons \cite{Bern:2010ue,Carrasco:2011mn,Bern:2011rj,Bern:2012uf,Boels:2012ew,Carrasco:2012ca,Bjerrum-Bohr:2013iza,Bern:2013uka,Bern:2014sna}, 
or only matter \cite{Nohle:2013bfa,Chiodaroli:2013upa,Johansson:2014zca}
(see also refs.~\cite{Mafra:2011kj,Mafra:2014gja,Mafra:2015mja,He:2015wgf}
for a string-theoretic approach).
The Cachazo-He-Yuan (CHY) representation for gauge theory amplitudes 
naturally encodes color-kinematic duality \cite{Cachazo:2013iea}.
In ref.~\cite{Johansson:2015oia}, 
Johansson and Ochirov established 
color-kinematic duality for amplitudes $\cA_{n,k}$ 
with $k$ pairs of quarks and $n-2k$ gluons
through explicit calculations for $n \le 8$.
For certain low-multiplicity amplitudes 
(specifically, $\cA_{4,1}$, $\cA_{5,2}$, and $\cA_{6,3}$)
they found that the numerators derived from Feynman rules
automatically satisfy the kinematic Jacobi relations. 
For higher-multiplicity amplitudes, 
a generalized gauge transformation is required 
to bring the kinematic numerators into a form 
that is manifestly color-kinematic dual.
\para

As discussed in the introduction,
color-kinematic duality implies new relations among 
color-ordered amplitudes for tree-level $n$-gluon amplitudes
\cite{Bern:2008qj}.
These were subsequently proven using string theory 
\cite{BjerrumBohr:2009rd,Stieberger:2009hq} 
and on-shell recursion \cite{Feng:2010my,Chen:2011jxa}, 
and necessarily hold for massless CHY 
amplitudes \cite{Cachazo:2012uq,Cachazo:2013iaa}. 
BCJ relations have also been established for amplitudes  
containing gluons and a pair of massive scalars 
in the fundamental representation by expressing 
these amplitudes in a CHY representation \cite{Naculich:2014naa}.
See ref.~\cite{Weinzierl:2014ava} for earlier work on a CHY 
representation for amplitudes containing gluons and 
massless quark-antiquark pairs. 
\para

Also as noted earlier,
Johansson and Ochirov derived relations 
among the primitive amplitudes in the Melia basis 
for $\cA_{n,k}$ that follow from color-kinematic duality;
these are the $k \ge 1$ analogs of the BCJ 
relations \cite{Johansson:2015oia}. 
They proceeded by expressing the Melia primitive amplitudes 
as linear combinations of $n_i/d_i$, 
and then imposing the Jacobi relations on the kinematic numerators.
Inverting these equations, 
they obtained equations for a subset of the $n_i$ 
in terms of the primitive amplitudes,
together with (for $n>2k$) equations 
among the primitive amplitudes.
Among their findings are that there are no relations among 
the primitive amplitudes for amplitudes containing no gluons ($n=2k$).
For amplitudes containing gluons, 
they found that the primitive amplitudes obey relations such as
\be
\sum_{b=3}^{n} \left( k_n \cdot k_1 
+ \sum_{c=3}^{b-1} k_n \cdot k_{\sigma(c)}  \right) 
A (1, \sigma(3), \cdots, \sigma(b-1), n, \sigma(b), \cdots, \sigma(n-1),2)  ~=~0
\label{bcj}
\ee
where $n$ denotes a gluon,    
and $\sigma$ is a permutation of $\{3, \cdots, n-1\} $.
\Eqn{bcj} has exactly the same form 
(when expressed in terms of invariants 
$k_a \cdot k_b$ where $k_a$ is the momentum of a gluon)
as one of the fundamental BCJ 
relations \cite{BjerrumBohr:2009rd,Feng:2010my,Sondergaard:2011iv} 
for an $n$-gluon amplitude.
Johansson and Ochirov found that relations of the form (\ref{bcj})
are satisfied when $n$ is replaced by the label of 
any of the other gluons in the amplitude, 
but not when $n$ is replaced by the label of a 
fundamental or antifundamental particle. 
Thus, the relations that hold for $k \ge 2$ are a proper subset
of the BCJ relations for $k=0$ or $k=1$.
These relations were obtained by Johansson and Ochirov
from explicit calculations for $n \le 8$ and for $n=9$, $k=4$,
and were proved for all $n$ using BCFW on-shell recursion 
by de la Cruz, Kniss, and Weinzierl \cite{delaCruz:2015dpa}.
The latter authors also presented a 
CHY representation for these amplitudes \cite{delaCruz:2015raa} 
(see also ref.~\cite{He:2016dol}).
\para

We will see in the next two sections that these results have a 
natural explanation in terms of the color-factor symmetry
possessed by the amplitude.
There is a set of color-factor shifts associated 
with each external gluon in the amplitude
and these give rise to the corresponding BCJ relations.
The absence of color-factor symmetry for amplitudes
with no gluons explains the absence of relations
among the Melia primitive amplitudes.
\para

For amplitudes containing only gluons, 
or amplitudes with gluons and one pair of fundamentals,  
the fundamental BCJ relations allow one 
to express the $(n-2)!$ amplitudes of the Kleiss-Kuijf basis 
in terms of a smaller basis of $(n-3)!$ amplitudes \cite{Bern:2008qj}.
Johansson and Ochirov found that 
for $k \ge 2$, the BCJ relations allow one to express the 
$(n-2)!/k!$ amplitudes of the Melia basis
in terms of a reduced basis of $(n-3)! (2k-2)/k!$ amplitudes \cite{Johansson:2015oia}.
The difference between 
$(n-2)!/k!$ and $(n-3)! (2k-2)/k!$ 
is precisely equal to $(n-2k) (n-3)!/k!$,
which as we will see is the dimension of 
the color-factor group for $k\ge 2$.

\section{Color-factor shifts}
\setcounter{equation}{0}
\label{sec:cfs}

In ref.~\cite{Brown:2016mrh}, we introduced the color-factor symmetry,
and proved that gauge-theory amplitudes containing gluons 
and massless or massive particles 
in an arbitrary representation of the gauge group
and with arbitrary spin $ \le 1$ are invariant under certain 
momentum-dependent shifts of the color factors.
In this section, we review the definition of these shifts,
and determine how they act on the Johansson-Ochirov color 
factors described in sec.~\ref{sec:proper}.
\para

Associated with each external gluon in the amplitude
is a set of symmetries that act as momentum-dependent 
shifts of the color factors appearing in the 
cubic decomposition (\ref{cubicdecomp}). 
Consider a tree-level color factor $c_i$ for an amplitude 
with an external gluon $a$.
The gluon leg divides the diagram in two at its point of attachment.  
Denote by $\Sai$ the subset of the remaining legs on one side
of this point;  it does not matter which side.
The action of the shift $\delta_a c_i$ is constrained by two requirements:
(I) that it preserve all the Jacobi relations satisfied by $c_i$,
and 
(II) that it satisfy 
\be
\delta_a c_i  ~\propto~  \sum_{c\in \Sai } k_a \cdot k_c   
\label{colorfactorshift}
\ee
where all momenta are outgoing.
(Choosing to sum over the complement of $\Sai$ gives the same result 
up to sign due to momentum conservation.)
In particular, if gluon $a$ is attached to an external leg $b$ with
momentum $k_b$, the shift is proportional to $k_a \cdot k_b$.
\para

Let $\{ c_i \}$ be the set of $n$-point color factors,
and consider the subset of them obtained from a 
fixed $(n-1)$-point cubic diagram by
attaching gluon $a$ in all possible ways.
One of these has gluon $a$ attached to external leg 1 
of the $(n-1)$-point diagram;
define its shift to be $\alpha\, k_a \cdot k_1$.
Then the conditions (I) and (II) above uniquely fix
the coefficients of the shifts of all the other 
color factors in this subset.
For example, in the five-point example discussed in the previous section
(see fig. 2), 
we define the color-factor shift associated with gluon 5 
(the only external gluon in the amplitude)  
to act on $c_1$ as
\be 
\delta_5 c_1 \equiv \alpha~ k_5 \cdot k_1 \,.
\label{fivepointshiftdef}
\ee 
Then using \eqn{fivepointjacobi},
we find the shifts of the other four color factors to be 
\be
\delta_5 c_2 ~=~ -\alpha~ k_5 \cdot k_2 \,, 
\qquad
\delta_5 c_3 ~=~ \alpha~ k_5 \cdot k_3 \,, 
\qquad
\delta_5 c_4 ~=~ -\alpha~ k_5 \cdot k_4 \,, 
\qquad
\delta_5 c_5 ~=~ -\alpha~ k_5 \cdot (k_1 + k_2)  \,.
\qquad
\label{fivepointshift}
\ee
The shifts of the JO color factors (\ref{fivepointJO})
associated with this amplitude are then 
\be
\delta_5 C_{15342} ~=~  \alpha~ k_1 \cdot k_5
\,,
\qquad
\delta_5 C_{13542} ~=~ \alpha~ (k_1+k_3) \cdot k_5
\,,
\qquad
\delta_5 C_{13452} ~=~ \alpha~ (k_1+k_3+k_4) \cdot k_5
\qquad
\ee
where a clear pattern emerges:  
the shift $\delta_a C_{\cdots a \cdots}$ depends on the
sum of momenta of the particles whose labels appear to the left
of $a$ in $C_{\cdots a \cdots}$.
\para

Now consider a general amplitude $\cA_{n,k}$ with at least one gluon.  
There is a set of color-factor symmetries 
for each of the $n-2k$ gluons, but to simplify the 
presentation, we will focus on the shift associated with gluon $n$.
Consider the set of JO color factors $C_{1n\sigma 2}$
where $\sigma$ is a permutation of $\{3, \cdots, n-1\}$.
Since $\sigma$ is a permutation of $n-2k-1$ gluons,
$k$ $\barpsi$'s, and $k$ $\psi$'s,
where the $\barpsi$'s and $\psi$'s form a Dyck word of length $2k-2$, 
the number of allowed choices of $\sigma$ is $(n-3)!/k!$.
Each
$C_{1n\sigma 2}$ is a linear combination of cubic color factors $c_i$, 
in each of which gluon $n$ is attached to external leg 1,
and which consequently have a shift proportional to $k_n \cdot k_1$. 
We therefore define the shift of $C_{1n\sigma 2}$ associated with
gluon $n$ to be 
\be
\delta_n~ C_{1  n \sigma  2} ~\equiv~ \alpha_{n,\sigma}\,  k_n \cdot k_1  
\label{defineshift}
\ee
where $\alpha_{n,\sigma}$ are a set of $(n-3)!/k!$ independent 
arbitrary constants.
Given \eqn{defineshift}, the shifts $\delta_n c_i$ 
of all other color factors are then uniquely determined.
The proof of this is as follows.   
The JO color factors form an independent basis 
in terms of which all the color factors $c_i$ can be expressed.  
In particular, all color factors $c_i$ 
with gluon $n$ attached to external leg 1
can be expressed in terms of the JO color factors $C_{1 n \sigma 2}$,
and therefore their shifts under $\delta_n$ are determined 
by \eqn{defineshift}.
But we argued above that the shifts of all color factors
are fixed once we know the shifts of the color factors 
that have gluon $n$ attached to external leg 1.
\para

Therefore we have shown that
associated with each gluon is an $(n-3)!/k!$-parameter 
family of color-factor shifts.
Including all the gluons, 
the color-factor shifts form an abelian group of 
dimension $(n-2k) (n-3)!/k!$ for $k\ge 2$.
For $k=1$, the color-factor shift associated with one of the gluons 
is a linear combination of those 
associated with all the others \cite{Brown:2016mrh}
and thus in that case the dimension of the color-factor group 
is $(n-3) (n-3)!$.
\para

We now show that, given \eqn{defineshift}, 
the shifts of the rest of the JO color factors are particularly simple, viz.
\be
\delta_n~ C_{1 \sigma(3) \cdots \sigma(b-1) n \sigma(b) \cdots  \sigma(n-1) 2 } 
~=~ 
\alpha_{n,\sigma} \left( 
k_n \cdot k_1  +  \sum_{c=3}^{b-1} k_n \cdot k_{\sigma(c)} \right) \,,
\qquad b=3, \cdots, n \,.
\label{JOshift}
\ee
We note that \eqn{JOshift} has exactly the same form as the 
shifts of the half-ladder color factors in ref.~\cite{Brown:2016mrh}; 
thus with respect to the color-factor symmetry, 
the JO color factors are the precise analog of the half-ladder color factors.
To establish \eqn{JOshift},
we consider the ``commutator'' of JO color factors
\be
C_{1 \sigma(3) \cdots \sigma(b-1) [\sigma(b) n] \sigma(b+1) \cdots  \sigma(n-1) 2 } 
~\equiv~
C_{1 \sigma(3) \cdots \sigma(b) n \sigma(b+1) \cdots  \sigma(n-1) 2 } 
-C_{1 \sigma(3) \cdots \sigma(b-1) n \sigma(b) \cdots  \sigma(n-1) 2 } 
\ee
or more briefly $C_{\cdots [cn] \cdots}$, where we let $c= \sigma(b)$.
We can most transparently compute this commutator using the 
inspired graphical notation of ref.~\cite{Johansson:2015oia}.
\para

\begin{figure}
\begin{center}
\includegraphics[scale=1.0,trim=150 680 150 50,clip=true]{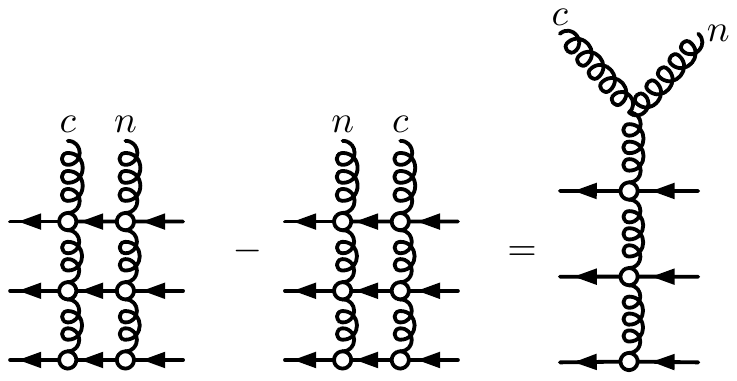}
\caption{Commutator $C_{\cdots [cn] \cdots}$ for gluon $g_c$.} 
\label{fig:u4}
\end{center}
\end{figure}

First, let $c$ label another gluon; 
then $C_{\cdots [cn] \cdots}$ is represented in fig.~\ref{fig:u4}.
The two diagrams on the l.h.s.~only fail to commute 
when the gluons are attached to the same line, 
in which case one can use 
$ [ T^{\ta}, T^{\tb} ]^{\ti}_{~\tj}=f_{\ta \tb \tc} ( T^{\tc} )^{\ti}_{~\tj} $
to give the diagram on the right.
Fig.~\ref{fig:u4} is the graphical depiction of the 
identity \cite{Johansson:2015oia}
\be
   \big[\Xi_{l}^\ta,\,\Xi_{l}^\tb \big]~=~f_{\ta \tb \tc}\,\Xi_{l}^\tc\,.
\ee
The diagram on the r.h.s.~of fig.~\ref{fig:u4}
is a linear combination of color factors in which gluon $n$ is attached to
external gluon $c$, 
and which therefore undergo a shift 
proportional to $k_n \cdot k_c$ under the color-factor symmetry
associated with gluon $n$.  
A little bit of thought shows that the coefficient 
of proportionality of the shift 
of the commutator is $\alpha_{n,\sigma}$ and therefore
\be
\delta_n 
C_{1 \sigma(3) \cdots \sigma(b-1) [\sigma(b) n] \sigma(b+1) \cdots  \sigma(n-1) 2 } 
~=~ \alpha_{n, \sigma} k_n \cdot k_{\sigma(b)}  \,.
\label{commutatorshift}
\ee
\begin{figure}[b]
\begin{center}
\includegraphics[scale=1.0,trim=100 700 100 50,clip=true]{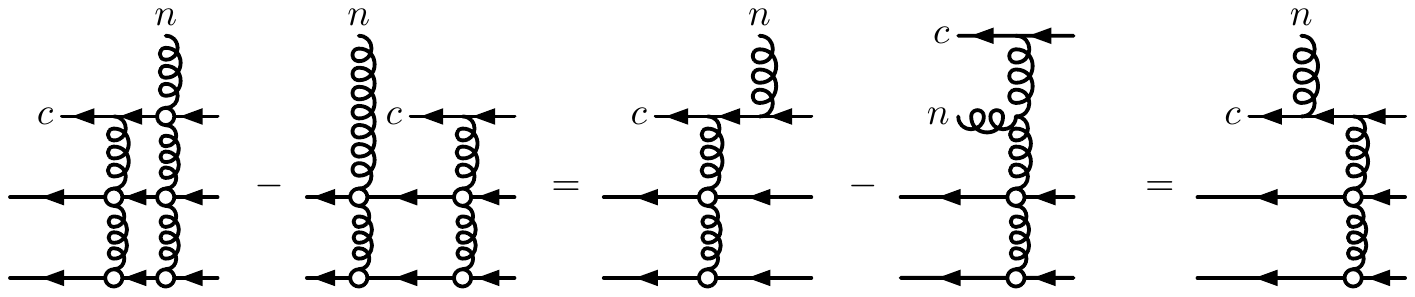}
\caption{Commutator $C_{\cdots [cn] \cdots}$ for $\barpsi_c$. }
\label{fig:u5}
\end{center}
\end{figure}

Next, let $c$ be the label of an antifundamental $\barpsi_c$;
the commutator $C_{\cdots [cn] \cdots}$
is then represented by fig.~\ref{fig:u5}.
The final diagram of the figure is a linear combination of 
color factors with gluon $n$ attached to $\barpsi_c$,
whose shifts are proportional to $k_n \cdot k_c$.
Again a bit of thought shows that the shift of this diagram
is given by  \eqn{commutatorshift}.
\para

\begin{figure}[t]
\begin{center}
\includegraphics[scale=1.0,trim=150 705 150 50,clip=true]{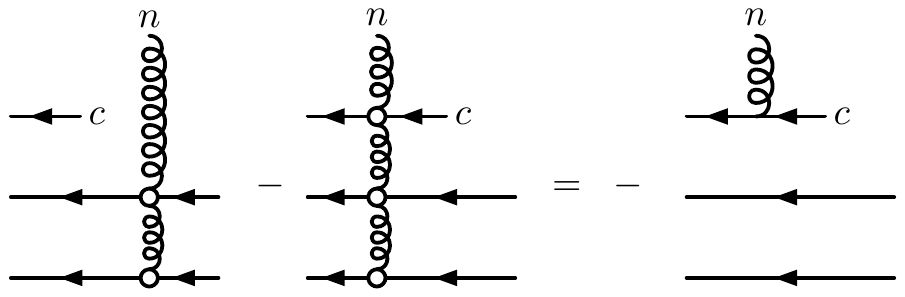}
\caption{Commutator $C_{\cdots [cn] \cdots}$ for $\psi_c$. }
\label{fig:u6}
\end{center}
\end{figure}

Finally, consider the case where $c$ labels a fundamental 
$\psi_c$;
the commutator $C_{\cdots [cn] \cdots}$ is shown in fig.~\ref{fig:u6}.
The r.h.s.~of the figure is a linear combination of color factors with 
gluon $n$ attached to $\psi_c$.
Taking account of the minus sign, 
the shift of the commutator is given by  \eqn{commutatorshift}.
\para

We can now apply \eqn{commutatorshift} recursively
starting with \eqn{defineshift} to obtain \eqn{JOshift}.   
The shift of the last JO factor is thus
\be
\delta_n 
C_{1 \sigma(3) \cdots  \sigma(n-1) n 2 } 
~=~ \alpha_{n,\sigma} 
\left(
k_n \cdot k_1  +  \sum_{c=3}^{n-1} k_n \cdot k_{\sigma(c)} 
\right)
~=~ - \alpha_{n,\sigma}  k_n \cdot k_2
\ee
which is consistent with the fact that 
$ C_{1 \sigma(3) \cdots  \sigma(n-1) n 2 } $
represents a linear combination of color factors in which
gluon $n$ is attached to $\psi_2$.
\para

In the next section we will use \eqn{JOshift} in the Melia-Johansson-Ochirov
decomposition to obtain the BCJ relations for $\cA_{n,k}$.

\section{BCJ relations from color-factor symmetry}
\setcounter{equation}{0}
\label{sec:bcjcfs}

In this section, we establish that the BCJ relations obtained by
Johansson and Ochirov for $\cA_{n,k}$ 
are a direct consequence of the color-factor symmetry
of the amplitude. 
We also show that the kinematic numerators for $\cA_{n,k}$ obey 
a set of gauge-invariant constraints that are less stringent
than the kinematic Jacobi relations, 
but which follow from the color-factor symmetry 
and are therefore sufficient to imply the BCJ relations.
\para

It was shown in ref.~\cite{Brown:2016mrh} that gauge-theory amplitudes 
with gluons as well as massless or massive  particles
in an arbitrary representation of the gauge group and arbitrary spin $\le 1$,
and therefore specifically $\cA_{n,k}$,
are invariant under the family of 
color-factor shifts described in sec.~\ref{sec:cfs}.
The proof of this uses the radiation vertex expansion 
of the amplitude \cite{Brown:1982xx}.  
A full description of the radiation vertex expansion 
and the proof of color-factor symmetry
is given in ref.~\cite{Brown:2016mrh}, 
but the basic strategy is as follows.
The radiation vertex expansion is a recursive approach
that  constructs an $n$-point amplitude by attaching a gluon
in all possible ways to all possible diagrams that contribute to
the $(n-1)$-point amplitude consisting of all the particles 
except for a chosen gluon $a$.
We may attach the gluon to an external leg, 
an internal line, or to one of the 
cubic $ggg$ or $\barpsi g \psi$ vertices (for $\psi$ a scalar or vector)
to make a quartic vertex.
Then all the contributions are reorganized into a sum over the legs 
of each of the vertices of each of the $(n-1)$-point diagrams.
The next step is to consider the action of the color-factor shift 
associated with gluon $a$ on the color factors appearing in
the radiation vertex expansion.
One proves that the sum over legs for each vertex
is invariant under the shift of color factors.
It immediately follows that the entire $n$-point amplitude
is invariant under the color-factor symmetry
associated with gluon $a$, 
\viz $\delta_a \cA_{n,k}=0$.
\para

Now consider an amplitude $\cA_{n,k}$ with at least one gluon $n$,
and consider the effect of the color-factor shift  $\delta_n$
associated with this gluon 
on the amplitude written in the Melia-Johansson-Ochirov
proper decomposition (\ref{mjo}).
Since the action of the shift on 
the Johansson-Ochirov color factors $C_{1\gamma 2}$
is given by \eqn{JOshift},
the shift acts on \eqn{mjo} as 
\begin{align}
&\delta_n \cA_{n,k} 
\nn\\
&~=~ 
\sum_{\sigma \in \text{Melia basis}} 
\alpha_{n,\sigma} 
\sum_{b=3}^{n} \left( k_n \cdot k_1 
+ \sum_{c=3}^{b-1} k_n \cdot k_{\sigma(c)}  \right) 
A (1,\sigma(3), \cdots, \sigma(b-1), n, \sigma(b), \cdots, \sigma(n-1),2)  \,.
\end{align}

and since $\alpha_{n,\sigma}$ are independent parameters, we conclude that
\be
\sum_{b=3}^{n} \left( k_n \cdot k_1 
+ \sum_{c=3}^{b-1} k_n \cdot k_{\sigma(c)}  \right) 
A (1, \sigma(3), \cdots, \sigma(b-1), n, \sigma(b), \cdots, \sigma(n-1),2)  ~=~0
\ee
precisely the fundamental BCJ relations 
obtained in ref.~\cite{Johansson:2015oia}.
The BCJ relations with $n$ replaced by another gluon $a$
follow from the invariance of the amplitude
under the color-factor shift associated with gluon $a$. 
There is no color-factor symmetry associated with gluonless amplitudes,
and therefore no BCJ relations among the Melia primitive amplitudes
are expected in that case, as was found in ref.~\cite{Johansson:2015oia}.
In sec.~\ref{sec:cfs}, we showed that 
the dimension of color-factor group is  $ (n-2k) (n-3)!/k!$
for $k \ge 2$, 
which reduces the number of independent primitives 
from the Melia basis of $(n-2)!/k!$ to $(2k-2) (n-3)!/k!$
as found in ref.~\cite{Johansson:2015oia}.
(For $k=1$, the color-factor group has dimension $(n-3) (n-3)!$
and thus reduces the number of independents from $(n-2)!$ to $(n-3)!$.)
\para

Although the BCJ relations (\ref{bcj}) were previously proven using 
on-shell BCFW recursion \cite{delaCruz:2015raa},
the proof in this paper based on color-factor symmetry
reveals a close connection between the BCJ relations 
and the symmetries of the Lagrangian formulation of gauge theory.
This connection is made explicit 
in the radiation vertex expansion proof of 
color-factor symmetry given in ref.~\cite{Brown:2016mrh}, 
and summarized in the discussion section of that paper,
which we briefly recap here. 
The variation of the amplitude 
under the color-factor shift 
associated with gluon $a$ 
can be separated into contributions
that are constant, linear, and quadratic 
in the gluon momentum $k_a$.
The $\cO(k_a^0)$ term 
is proportional to $\sum_r \ve_a \cdot K_r$,
where $K_r$ are the momenta flowing out of each vertex. 
This vanishes by $\ve_a \cdot k_a = 0$
together with momentum conservation 
(a result of spacetime translation invariance of the Lagrangian).
The $\cO(k_a^1)$ term 
of the variation of the amplitude 
is given by a sum of
angular momentum generators $J_r^{\alpha\beta}$,
which act as a first-order Lorentz transformation 
on the relevant vertex factors.
These terms vanish by Lorentz invariance of the Lagrangian.
The vanishing of the $\cO(k_a^2)$ term
of the variation of the amplitude is more subtle,
but relies on Poincar\'e invariance
together with Yang-Mills gauge symmetry.
Thus, the vanishing of the variation of the amplitude
under the color-factor shift
(and therefore the BCJ relations) 
is closely tied to 
(if not quite a direct consequence of)
the gauge and Poincar\'e symmetries
of the gauge theory.
\para

We also introduced in ref.~\cite{Brown:2016mrh} the 
cubic vertex expansion of an amplitude containing at least one gluon. 
(This is related to, but distinct from, the radiation vertex expansion.)
Consider the set of cubic diagrams $I$ that contribute to the $(n-1)$-point
amplitude of all the particles in $\cA_{n,k}$ except for gluon $a$.
For any $a$, 
the amplitude $\cA_{n,k}$ can be written as a triple sum
over the legs $r$ of the vertices $v$ of the cubic diagrams $I$:
\be
\cA_{n,k} ~=~ 
\sum_I  
\sum_v  {1 \over \prod_{s=1}^3 d_{(a,I,v,s)}  }
\sumr 
{ c_{(a,I,v,r)} n_{(a,I,v,r)} \over 2 k_a \cdot K_{(a,I,v,r)}   } \,.
\label{cubicvertexexpansion}
\ee
Here  
$d_{(a,I,v,r)}$ is the product of propagators\footnote{If
leg $r$ is external, then $d_{(a,I,v,r)}=1$.}
that branch off from leg $r$ of vertex $v$ of diagram $I$,
$K_{(a,I,v,r)}$ is the momentum flowing out of that leg,
$c_{(a,I,v,r)}$ is the color factor of the $n$-point diagram 
obtained by attaching gluon $a$ to leg $r$ of vertex $v$ of diagram $I$,
and $n_{(a,I,v,r)}$ is the $n$-point kinematic numerator 
associated with that color factor.
As a concrete example, consider the five-point amplitude 
discussed in sec.~\ref{sec:proper}.
Using the identity 
\be
{1 \over s_{12} s_{34} }
~=~
{1 \over 2 s_{34} (-k_1-k_2)  \cdot k_5 }
~+~
{1 \over 2 s_{12} (k_1+k_2)  \cdot k_5 }
\ee
it is straightforward to write 
\eqns{fivepointcubicdecomp}{fivepointcolorfactors}
as 
\be
\cA_{5,2}
~=~
{1 \over s_{34} }
\left[    
 {c_1 n_1 \over 2 k_1 \cdot k_5}
+{c_2 n_2 \over 2 k_2 \cdot k_5}
+{c_5 n_5 \over 2(-k_1-k_2)  \cdot k_5}
\right]
+
{1 \over s_{12} }
\left[    
{c_3 n_3 \over 2 k_3 \cdot k_5}
+{c_4 n_4 \over 2 k_4 \cdot k_5}
+{c_5 n_5 \over 2 (k_1+k_2)  \cdot k_5}
\right]
\ee
which is precisely of the form (\ref{cubicvertexexpansion}).
\para

The color factors appearing in \eqn{cubicvertexexpansion} obey 
$\delta_a \, c_{(a,I,v,r)} ~=~ \alpha_{(a,I,v)} \,k_a \cdot K_{(a,I,v,r)}$
under the shift associated with gluon $a$.
Since $\delta_a \cA_{n,k}=0$,
we may conclude from the cubic vertex expansion
(\ref{cubicvertexexpansion}) that
\be
\sum_I  \sum_v
{\alpha_{(a,I,v)}  \over \prod_{s=1}^3 d_{(a,I,v,s)}  }
\sumr 
n_{(a,I,v,r)}  ~ = ~ 0 \,.
\label{sumoverdelta}
\ee
Because the $\alpha_{(a,I,v)} $ are not independent\footnote{The set 
of $\alpha_{(a,I,v)}$ for all the vertices of a given diagram $I$
are equal (up to signs) because any two adjacent vertices 
share a common color factor.
Moreover $\alpha_{(a,I,v)}$ must respect the Jacobi relations 
among the color factors for different diagrams $I$.}
we may not draw the more stringent conclusion that 
$\sumr n_{(a,I,v,r)} =0$ for each vertex.
For the five-point amplitude considered above, \eqn{sumoverdelta}
yields only one constraint
\be
0 
~=~ 
 \delta_5 \cA_{5,2}
~=~ {\alpha \over 2}
\left[ {n_1  - n_2  + n_5 \over s_{34} }
~+~
{n_3  - n_4  - n_5 \over s_{12} }
\right]
\ee
rather than the two kinematic Jacobi 
relations\footnote{For this five-point example 
with spin one-half fundamentals,
Johansson and Ochirov found that the numerators derived from the 
Feynman rules automatically satisfy the kinematic Jacobi relations.}
$n_1  - n_2  + n_5 =0$ and $n_3  - n_4  - n_5 =0$.
In general, color-kinematic duality states that 
a generalized gauge transformation exists such that
the numerators {\it in that gauge} obey the 
kinematic Jacobi relations.
The relation (\ref{sumoverdelta}), however,
holds for the kinematic numerators in any gauge,
since it is invariant under generalized gauge transformations,
as can be seen from its derivation.

\section{Conclusions}
\setcounter{equation}{0}
\label{sec:concl}

In this paper, we have shown that 
BCJ relations \cite{Johansson:2015oia}
among the Melia-basis primitive amplitudes of $\cA_{n,k}$ 
with $n-2k$ gluons and $k$ pairs of particles in the fundamental 
(or other) representation of the gauge group 
follow directly from the invariance of
$\cA_{n,k}$ 
under a set of color-factor shifts.
We have also derived 
as a consequence of this symmetry
a set of gauge-invariant constraints
on the kinematic numerators of $\cA_{n,k}$.
\para

The tree-level color-factor symmetry has been proven for a  
wide class of gauge-theory amplitudes, including those with
massless or massive particles with gauge-theory couplings 
in arbitrary representations of the gauge group and 
arbitrary spin $\le 1$
\cite{Brown:2016mrh}.
This is connected to the radiation symmetry coming out of
theorems on photon radiation zeros in refs.~\cite{Brown:1982xx,Brown:1983pn}.
The color-factor symmetry also applies to theories with gauge bosons 
that become massive through spontaneous symmetry breaking 
(\eg see refs.~\cite{Naculich:2015zha,Naculich:2015coa,Chiodaroli:2015rdg}).
The only particles in the amplitude that need be massless are the
gluons (or photons) with which the color-factor symmetries are associated.
Thus it applies to standard-model gauge theory amplitudes
as well as to many extensions thereof.
\para

BCJ relations are constraints among 
gauge-invariant primitive amplitudes,
the coefficients in a proper decomposition 
of the gauge-theory amplitude.
Such decompositions have been identified for tree-level
and one-loop $n$-gluon amplitudes in ref.~\cite{DelDuca:1999rs}
and for the tree-level amplitudes considered in this paper in
refs.~\cite{Melia:2013bta,Melia:2013epa,Melia:2015ika,Johansson:2015oia}.
Once proper decompositions for more general amplitudes have been identified,
relations among their primitive amplitudes
will follow as a consequence of color-factor symmetry.
\para

Finally, it was shown in ref.~\cite{Brown:2016mrh}
that one-loop amplitudes that have color-kinematic-dual representations
are invariant under a loop-level generalization 
of the color-factor symmetry,
although a proof based on Lagrangian methods is still lacking.
One can legitimately hope that color-factor symmetry 
will soon lead to many new insights into gauge theories
in general and color-kinematic duality in particular.

\section*{Acknowledgments}
We would like to thank H.~Johansson and A.~Ochirov for sharing 
with us their files for producing figures.
This material is based upon work supported by the
National Science Foundation
under Grants Nos.~PHY14-16123 and PFI:BIC 1318206.
RWB is also supported by funds
made available through a CWRU Institute Professorship Chair.
SGN gratefully acknowledges sabbatical support
from the Simons Foundation (Grant No.~342554 to Stephen Naculich).
He would also like to thank the Michigan Center for Theoretical Physics
and the Physics Department of the University of Michigan 
for generous hospitality 
and for providing a welcoming and stimulating sabbatical environment.

\vfil\break


\end{document}